\documentclass[prl,reprint,showpacs,showkeys,aps]{revtex4-1}
\usepackage{amssymb,amsmath,amsthm,mathtools}
\usepackage{graphicx}
\usepackage{units}
\usepackage[usenames,dvipsnames,svgnames,table]{xcolor}

\newcommand{\pa}{\partial_1}
\newcommand{\pb}{\partial_2}
\newcommand{\pt}{\partial_t}                                     
\newcommand{\expect}[1]{\left< #1 \right>}                                     
\newcommand{\braket}[2]{\left< #1 \vphantom{#2} \right| \left. #2 \vphantom{#1} \right>}
\newcommand{\matrixel}[3]{\left< #1 \vphantom{#2#3} \right| #2 \left| #3 \vphantom{#1#2} \right>}

\begin{document}

\title{A single-electron approach for many-electron dynamics in high-order 
harmonic generation}
\author{Axel Schild}
\affiliation{Max-Planck-Institut f\"ur Mikrostrukturphysik}
\author{E.K.U. Gross}
\affiliation{Max-Planck-Institut f\"ur Mikrostrukturphysik}

\date{\today}

\begin{abstract}
  We present a novel ab-initio single-electron approach to correlated 
  electron dynamics in strong laser fields. 
  By writing the electronic wavefunction as a product of a marginal 
  one-electron wavefunction and a conditional wavefunction, we show that 
  the exact harmonic spectrum can be obtained from a single-electron 
  Schr\"odinger equation.
  To obtain the one-electron potential in practice, we propose an adiabatic 
  approximation, i.e.\ a potential is generated that depends only on the 
  position of one electron.
  This potential, together with the laser interaction, is then used to obtain 
  the dynamics of the system.
  For a model Helium atom in a laser field, we show that by using our approach, 
  the high-order harmonic generation spectrum can be obtained to a good 
  approximation.
\end{abstract}
%

\maketitle

One of the most interesting phenomena occurring in the interaction of 
molecules with strong laser fields is the process of high-order harmonic
generation (HHG) \cite{mcpherson87,ferray88}: a molecule exposed to
an infrared driving laser radiates light with a frequency being an integer 
multiple of the driving frequency, with significant intensities.
As the electron dynamics responsible for the light emission is fast, the 
resulting spectra can be used as an attosecond probe of the dynamics of 
the molecule.\cite{itatani04,krausz09,smirnova09,haessler10,haessler11,
salieres12}

The process of HHG is most easily understood in the 3-step model 
\cite{schafer93,corkum93,lewenstein94}, which is a semi-classical 
single-electron picture: During the laser cycle, an electron first 
tunnels out of the bound-state region of the molecule. 
The laser field then accelerates it away and, when the field switches sign, 
back to the molecule.
Finally, the collision of the electron with the molecule leads to the measured 
emission of high-order harmonics.
In practice, a single-active electron approximation \cite{kulander87} is 
often used to calculate spectra:
For a suitable model potential, a one-electron Schr\"odinger equation is solved 
and the spectrum is calculated.
Finding an appropriate potential is a challenge \cite{abusamha10,awasthi10} 
and the correct incorporation of many-electron effects poses an obstacle.

Clearly, the merit of the single-active electron approximation is that it is 
only necessary to solve a single-particle Schr\"odinger equation.
In this article, we want to retain this crucial advantage, but ask the question:
Is it possible to obtain the relevant dynamical data to calculate the 
HHG spectra from a single-electron calculation not only approximately, 
but exactly?
The answer is yes, and the way how we answer this question suggests a strategy 
to obtain the corresponding potential in practice.

Our approach is inspired by the exact factorization (EF) of a 
time-dependent molecular 
wavefunction into a marginal nuclear and a conditional electronic 
wavefunction \cite{cederbaum08,abedi10,abedi12,abedi13,agostini13,abedi14,agostini14,
agostini15a,min15,scherrer15,suzuki15,agostini15b,khosravi15}, see also the 
time-independent case \cite{hunter75,hunter81,gidopoulos14,chiang14,min14,
requist15}.
The EF is normally used to separate the molecular wavefunction into a nuclear 
and an electronic wavefunction.
However, the EF was also used to describe electron localization in H$_2^+$ by 
strong lasers \cite{suzuki14} (by formally reversing the role of nuclei and 
electrons), to introduce time into the stationary Schr\"odinger 
equation \cite{arce12}, and to factor identical 
particles.\cite{hunter86,buijse89}

We build upon that latter approach and apply the EF to the time-dependent 
electronic wavefunction to obtain an exact single-electron equation.
Complementary to this, we use an expansion in adiabatic states (ASE), the 
analogue of the Born-Huang expansion \cite{born54}, as a tool for analysis.
Ultimately, we apply an adiabatic approximation reminiscent of the 
Born-Oppenheimer approximation \cite{born27} and check its validity.

For this purpose, we consider a system of 2 electrons at positions $r_1, r_2$ 
in the external field of the laser and of the (clamped) nuclei, and, for 
simplicity, spin is not considered explicitly.
The interaction of the electrons with the laser is treated in the dipole 
approximation, so that the system is described by the Schr\"odinger equation
\begin{equation}
 i \pt \psi = \left( -\frac{\pa^2}{2} -\frac{\pb^2}{2} + V(r_1,r_2) + (r_1+r_2) F(t) \right) \psi
  \label{eq:sys}
\end{equation}
for $\psi(r_1,r_2,t)$. Here, $\partial_i$ is the gradient w.r.t.\ $r_i$ and $V(r_1,r_2)$ includes the 
electron-electron repulsion and the interaction of the electrons with the 
clamped nuclei.
Atomic units are used throughout the article.

The restriction to 2 electrons is done only for convenience of notation, but 
the method described below can be applied to any number of electrons. 
Typically, $r_1$ are the coordinates of one electron and $r_2$ are those of 
the other electrons, but $r_1$ may also contain coordinates of more than one 
electron.
This is important if two- or many-electron observables are of interest, like 
double ionization, etc.

In the EF, it is used that any joint probability distribution function (PDF) 
depending on several variables can be written as a product of a marginal PDF 
and a conditional PDF. 
The marginal PDF depends on a subset of the variables, and the conditional 
PDF depends on the other ones and conditionally on the subset.
Thus, the time-dependent wavefunction is written as a product
\begin{align}
 \psi(r_1,r_2,t) &= \chi(r_1,t) \phi(r_2,t|r_1) 
  \label{eq:eef}
\end{align}
with a marginal wavefunction
\begin{align}
 \chi(r_1,t) &= e^{-i S(r_1,t)} \sqrt{\rho(r_1,t)} 
  \label{eq:chief}
\end{align}
and a conditional wavefunction
\begin{align}
 \phi_j(r_2,t|r_1) &= \psi(r_1,r_2,t) / \chi(r_1,t).
  \label{eq:phief}
\end{align}
The one-electron density $\rho(r_1,t) = \braket{\psi}{\psi}_2$ is the marginal 
PDF which gives the probability of finding an electron at $r_1$ independent of 
the other electron.
Here, $\braket{\cdot}{\cdot}_i$ indicates integration over $r_i$,
For $\phi_j$ the partial normalization condition 
$\braket{\phi_j}{\phi_j}_2 = 1$ is fulfilled for all $r_1, t$.
Thus, $|\phi_j(r_2,t|r_1)|^2$ is a conditional PDF, i.e., given an electron 
at $r_1$, it gives the probability of finding the other electron at $r_2$.

The equation of motion for the marginal wavefunction is \cite{abedi12}
\begin{align}
  i \pt \chi &= \left( -\frac{(\pa+A(r_1,t))^2}{2} + \epsilon(r_1,t) \right) \chi
   \label{eq:chieom}
\end{align}
with the vector potential
\begin{align}
  A(r_1,t) = \operatorname{Im} \braket{\phi}{\pa \phi}_2 
   \label{eq:pap}
\end{align}
and scalar potential
\begin{multline}
  \epsilon(r_1,t) = \matrixel{\phi}{-\frac{\pb^2}{2} + V(r_1,r_2)}{\phi}_2 + \operatorname{Im} \braket{\phi}{\pt \phi}_2 \\ + \frac{\braket{\pa\phi}{\pa\phi}_2-A(r_1,t)^2}{2} + \left(r_1 + \matrixel{\phi}{r_2}{\phi}_2 \right) F(t).
   \label{eq:pep}
\end{multline}
The latter consists of a part similar to a Born-Oppenheimer potential 
(albeit with the exact, time-dependent conditional wavefunction $\phi$), a 
gauge-dependent part, a part that is related to a Fubini-Study 
metric\cite{berry89}, and the laser interaction.
The gauge transformation
\begin{align}
 \tilde{\chi} = e^{-\tilde{S}(r_1,t)} \chi, ~~~ \tilde{\phi} = e^{\tilde{S}(r_1,t)} \phi \label{eq:gauge} 
\end{align}
leaves the total wavefunction and  the equations of motion unchanged but 
transforms the vector and scalar potential as $\tilde{A} = A + \pa \tilde{S}$ 
and $\tilde{\epsilon} = \epsilon + \pt \tilde{S}$, respectively.
The equation of motion for $\phi$ is not being used in the present context and 
can be found, for example, in \cite{abedi12}.

The EF formulation is the exact single-electron picture that we are looking for.
As was shown in \cite{abedi10}, $\chi$ and $\phi$ are unique up to the gauge 
transformation \eqref{eq:gauge}, and $\chi$ gives the correct one-electron 
density and one-electron current density and, hence, yields the exact 
HHG spectrum.
The EF may thus be seen as the, so far missing, formal justification of the 
single-active electron approach. \cite{rohringer06}

In the ASE, the wavefunction is written as
\begin{align}
 \psi(r_1,r_2,t) &= \sum_{j=0}^{\infty} \chi_j^{\rm ASE}(r_1,t) \phi_j^{\rm ASE}(r_2|r_1),
  \label{eq:ase}
\end{align}
where the wavefunctions $\phi_j^{\rm ASE}(r_2|r_1)$ are obtained as stationary 
states for fixed
$r_1$,
\begin{align}
 \epsilon_j^{\rm ASE}(r_1) \phi_j^{\rm ASE} &= \left(-\frac{\pb^2}{2} + V(r_1,r_2) \right) \phi_j^{\rm ASE}. 
  \label{eq:phias}
\end{align}
Clearly, the wavefunctions $\phi_j^{\rm ASE}$ are analogous to the 
Born-Oppenheimer states in the electron-nuclear case.
Ansatz \eqref{eq:ase} leads to coupled equations of motion for the coefficients 
$\chi_j^{\rm ASE}$ \cite{malhado14}.
The analogue of the Born-Oppenheimer approximation for a selected state $k$ is 
neglect of the coupling terms, i.e.\
\begin{align}
 i \pt \chi^{\rm AE} &= \left( -\frac{\pa^2}{2} + \epsilon_k^{\rm ASE}(r_1) + V_L(r_1,t) \right) \chi^{\rm AE},
  \label{eq:chias}
\end{align}
with the laser interaction
\begin{align}
 V_L(r_1,t) &= \left(r_1 + \braket{\phi_k^{\rm ASE}}{r_2 \phi_k^{\rm ASE}}\right) F(t).
  \label{eq:a2} 
\end{align}

Equation \eqref{eq:chias} is our proposal for a single-electron approach to 
the electron dynamics in strong laser fields.
For $N$ electrons, we propose to find the stationary state for one electron 
fixed at position $r_1$ by solving \eqref{eq:phias} for a suitable state $k$.
Then, with the potential generated by varying $r_1$, the dynamics of the full 
$N$-electron system is replaced by the dynamics of the marginal single-electron 
system obtained from \eqref{eq:chias}.
We call the dynamics arising from \eqref{eq:chias} the adiabatic electron (AE) 
approximation.
The AE approximation takes into account the dynamics of the cation as well as 
the effect of the cation on the ionized electron.
In contrast to the single-active electron approximation, it gives a unique 
procedure for obtaining the single-electron potential for any molecular 
system.

As all electrons have the same mass, the AE approximation 
\eqref{eq:chias} is not as generically useful as the Born-Oppenheimer 
approximation.
Also, it breaks the anti-symmetry of the electronic 
wavefunction, albeit only w.r.t.\ the one electron at $r_1$.
For the electrons in the set $r_2$, the symmetry requirements are obeyed.
In situations like ionization and HHG, when one electron moves far away from 
the rest, the lack of anti-symmetrization of one electron with the wavefunction 
of the remaining cation is not expected to be a serious restriction for the 
description of the process.

To test the AE approximation, we study a numerically exactly 
solvable model of two electrons in an atom. 
We choose the model of \cite{shilin13}:
A helium atom is modeled by the soft-Coulomb interaction potentials
\begin{align}
 V(r_1,r_2) &= \frac{1}{\sqrt{(r_1-r_2)^2+c}} - \sum_{j=1}^2 \frac{2}{\sqrt{r_j^2+c}} 
  \label{eq:vpot}
\end{align}
where $r_1, r_2$ are now the coordinates of electron 1 and 2, each in one 
dimension.
The laser pulse is a 12-cycle pulse $F(t) = F_0 \cos(\omega t) f(t)$, where 
$\tau = 2 \pi / \omega$ is the laser period and $f(t)$ increases (decreases) 
linearly from 0 to 1 during the first (last) two cycles.
The laser frequency $\omega$ corresponds to a wavelength of \unit[530]{nm}, and 
the laser amplitude $F_0$ yields a maximum intensity of 
\unit[$6.88 \times 10^{14}$]{$W/cm^2$}.
The interaction parameter is \unit[$c = 0.55$]{$a_0^2$}.
For the calculations we used at least the spacial and temporal grid of 
\cite{shilin13}, with the same mask region (the wavefunction is absorbed if 
$|r_i| >$ \unit[70]{$a_0$}).
The static eigenvalue problems were solved by implicitly restarted Arnoldi 
methods \cite{lehoucq98} as implemented in SciPy \cite{scipy}, while the time 
propagation was performed with a split-operator method \cite{feit82}.

\begin{figure}[htbp]
  \centering
  \includegraphics[width=0.49\textwidth]{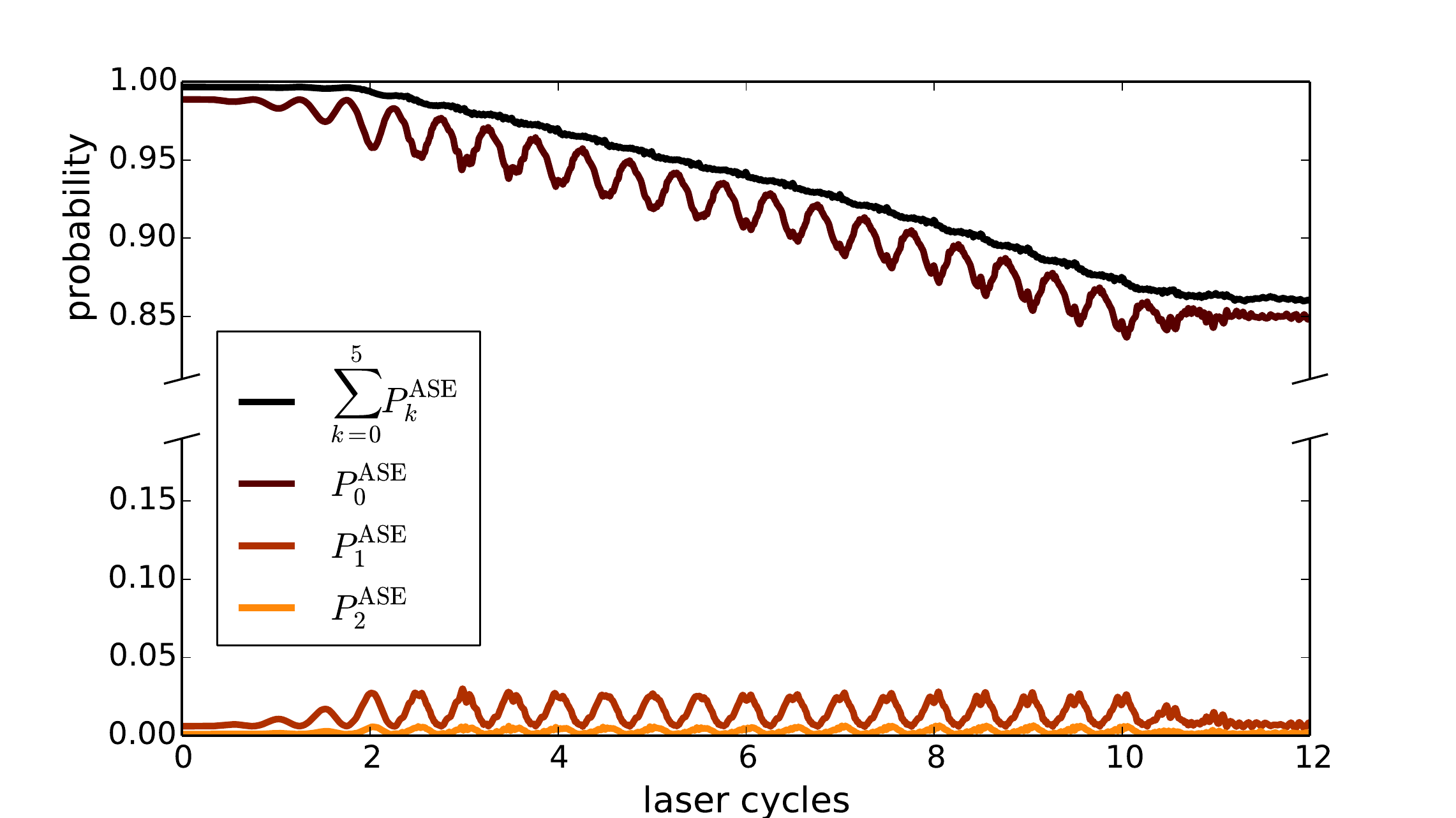}
  \caption{
  Occupation numbers $P_j^{\rm ASE}$ \eqref{eq:occ} of the adiabatic state 
  expansion \eqref{eq:ase} and partial sum for the lowest 6 states (visually 
  indistinguishable from the norm of the wavefunction).}
  \label{fig:expansion}
\end{figure}

First, we compute the occupation numbers
\begin{align}
 P_j^{\rm ASE}(t) = \braket{\chi_j^{\rm ASE}}{\chi_j^{\rm ASE}}_1
  \label{eq:occ}
\end{align}
of the ASE states from the exact wavefunction, shown in figure 
\ref{fig:expansion}.
In contrast to an expansion of the full wavefunction in terms of the 
eigenstates of the system, there are no ``symmetry selection rules'' in 
the ASE and all states can be occupied.
The wavefunction is mainly composed of the ASE ground state, with only small 
contributions from the first and second excited state and a periodic population 
transfer between ground and excited states with a period of half the laser 
period.
Consequently, an AE approximation to consider only the ground state for the 
dynamics should be applicable.

\begin{figure}[htbp]
  \centering
  \includegraphics[width=0.49\textwidth]{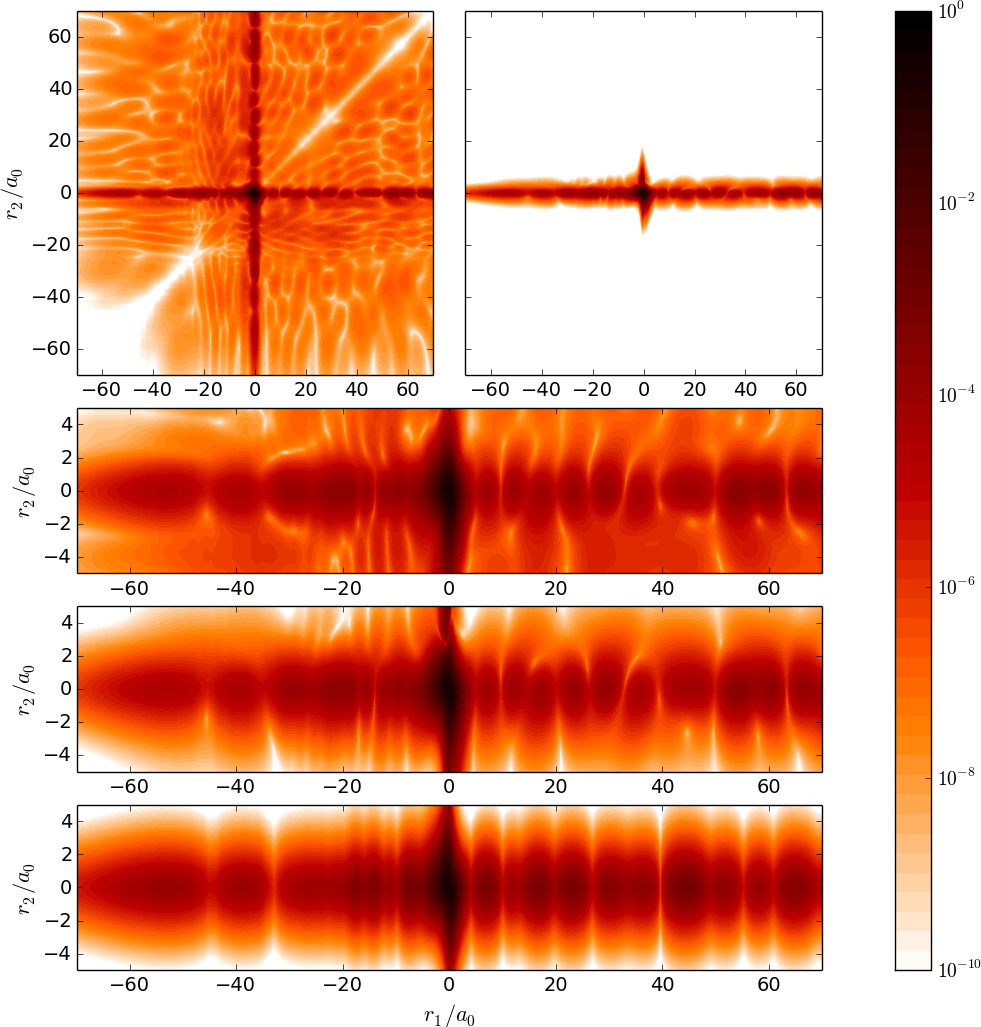}
  \caption{
  Top: Logarithmic plot of the exact density $|\psi(r_1,r_2,t)|^2$ for 
  $t = 6 \tau$ (after half of the pulse) and of the  density 
  $|\psi^{(3)}(r_1,r_2,t)|^2$ for the expansion of $\psi$ in the ASE, 
  \eqref{eq:ase}, including only the lowest 3 states, 
  $\psi^{(3)} = \sum_{k=0}^2 \chi_j^{\rm ASE}(r_1,t) \phi_j^{\rm ASE}(r_2|r_1)$. 
  Below: Blow up of $|\psi|^2$, $|\psi^{(3)}|^2 $, and the density of the 
  adiabatic electron approximation $|\psi^{\rm AE}(r_1,r_2,t)|^2$ for 
  $\psi^{\rm AE} = \chi^{\rm AE}(r_1,t) \phi_0^{\rm ASE}(r_2|r_1)$.}
  \label{fig:expansion_rho}
\end{figure}

The occupation numbers give only an incomplete picture:  
They do not reveal that only certain parts of the full wavefunction are 
represented well by a truncated expansion.
In figure \ref{fig:expansion_rho} we show a logarithmic plot of the density 
after half of the laser pulse is over (at a time where the field is maximal).
In the top row of the figure, the exact density and the density obtained from 
an ASE truncated after 3 states are shown.
Clearly, the symmetry properties of the wavefunction are lost and any 
two-electron observables, notably double ionization (the region where both 
$|r_1|$ and $|r_2|$ are large), cannot be described by the truncated 
wavefunction.
In such a case, a repeated factorization of the many-electron 
wavefunction, like described in \cite{cederbaum15}, may be useful to derive
computational methods.

The loss of the symmetry properties is by construction:
We allow $r_1$ to become as large as necessary, but along $r_2$ we compute the 
bound states.
The energetically lowest bound states are localized in the region of small $r_2$.
Hence, for describing large excursions along $r_2$ (which is necessary when we 
want to recover the correct symmetry of the wavefunction) we need highly 
excited states and continuum states.

However, for small $r_2$ the density is well-reproduced by the truncated 
expansion.
The two central panels of figure \ref{fig:expansion_rho} show a blow-up of the 
respective region.
Details of the exact density are already well reproduced by the first few 
states of the expansion and inclusion of a few further states would make the 
dynamics in this region almost indistinguishable from the exact one.
Hence, we conclude that observables related to a large excursion of only one 
electron can be described with a truncated expansion, as intuitively expected.

In the bottom panel of figure \ref{fig:expansion_rho}, a plot of the density at 
$t=6\tau$ is shown for a propagation in the AE approximation, i.e., obtained 
from solving \eqref{eq:chias} using the ground state $\phi_0$ of 
\eqref{eq:phias}.
The density is symmetric w.r.t.\ $r_2$ because $\phi_0$ has this property.
While there are discrepancies from the exact density, the qualitative structure 
is reproduced, even on this logarithmic scale.

\begin{figure}[htbp]
  \centering
  \includegraphics[width=0.49\textwidth]{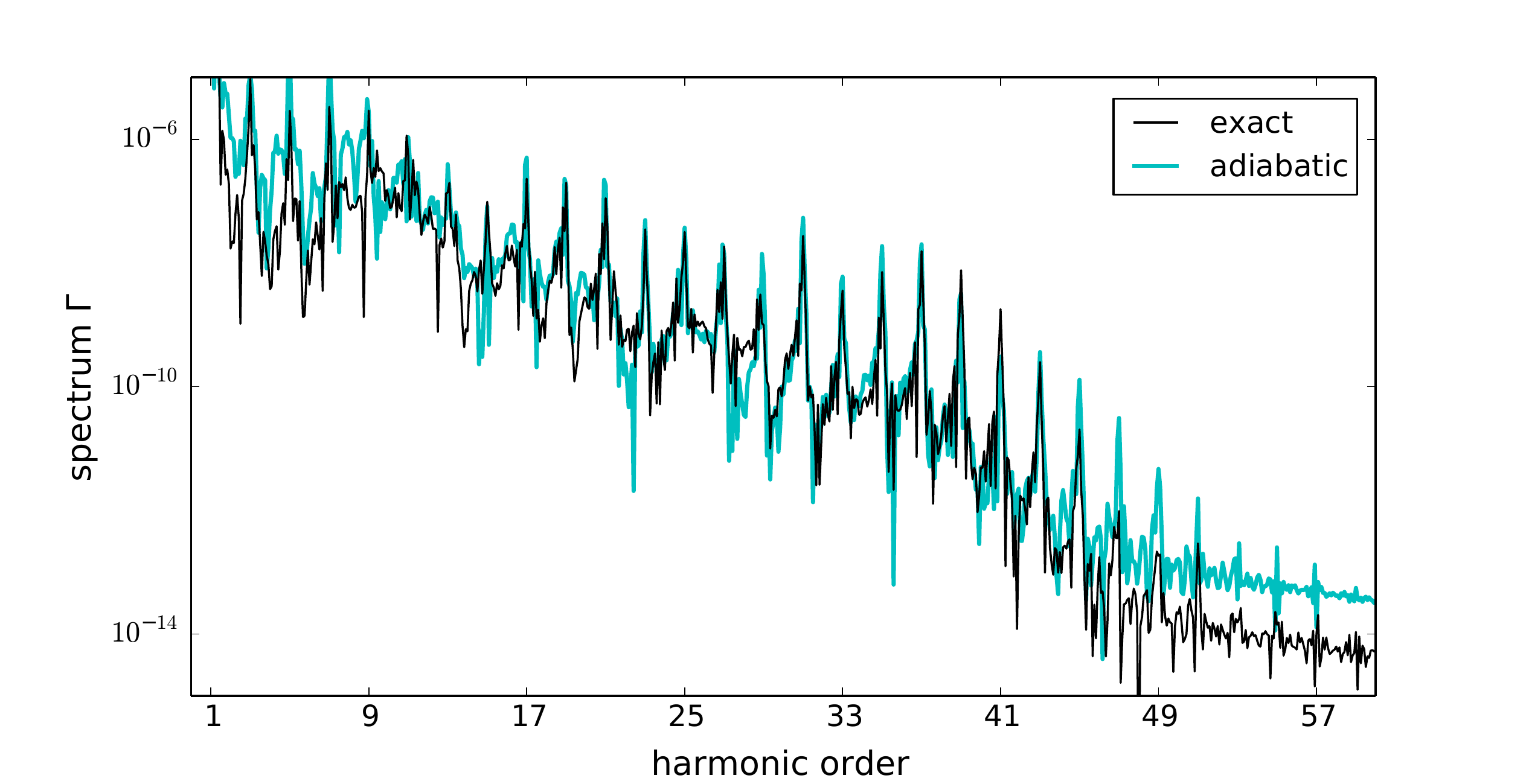}
  \caption{
  Logarithmic plots of the spectrum \eqref{eq:spectrum} for the exact dynamics and the dynamics obtained from the adiabatic electron approximation \eqref{eq:chias}, for the model of \cite{shilin13}.}
  \label{fig:spectrum_cpb}
\end{figure}

The spectrum obtained from the AE approximation is calculated as the Fourier 
transform of the dipole acceleration \cite{bandrauk09},
\begin{align}
 \Gamma(\omega) = \frac{2}{T^2 \omega^4} \left| \int\limits_0^T \pt^2 \expect{r_1 \rho}_1 e^{-i \omega t} dt \right|,
  \label{eq:spectrum}
\end{align}
and shown in figure \ref{fig:spectrum_cpb}.
We choose \eqref{eq:spectrum} to compute the spectrum because it only needs the 
directly accessible 
one-electron density $\rho(r_1,t)$, but in principle other methods can be used.
The factor 2 is the number of electrons in the system.

As can be seen from the figure, the overall features of the exact spectrum are 
well reproduced,
especially the frequency region from harmonic 9 to harmonic 43.
Only initially and close to the cutoff frequency the approximate spectrum 
deviates from the exact one.
The discrepancies can be explained by a comparison between the exact potential 
$\epsilon$ of \eqref{eq:pep} and the approximate potential 
$\epsilon_0^{\rm ASE}+V_L$ of \eqref{eq:chias}, for a gauge where the exact 
vector potential $A = 0$.
While they are almost indistinguishable in the bound-state region, for the 
region of the barrier that the electron has to tunnel through we find 
$\epsilon \ge \epsilon_0^{\rm ASE}+V_L$.
The difference is small but noticeable in the dynamics, because it leads to 
higher values of the density outside the core region.
In other words, during the time of maximum field amplitude the electron has a 
higher chance to tunnel out of the bound-state region than it should have.
Furthermore, there are spikes\cite{czub78,hunter81,chiang14} and 
steps\cite{elliot12,abedi13} in the potential.
They are signs of non-adiabatic effects, i.e.\ they show that restriction of 
the dynamics to only the ground state potential leads to errors in the dynamics.

\begin{figure}[htbp]
  \centering
  \includegraphics[width=0.49\textwidth]{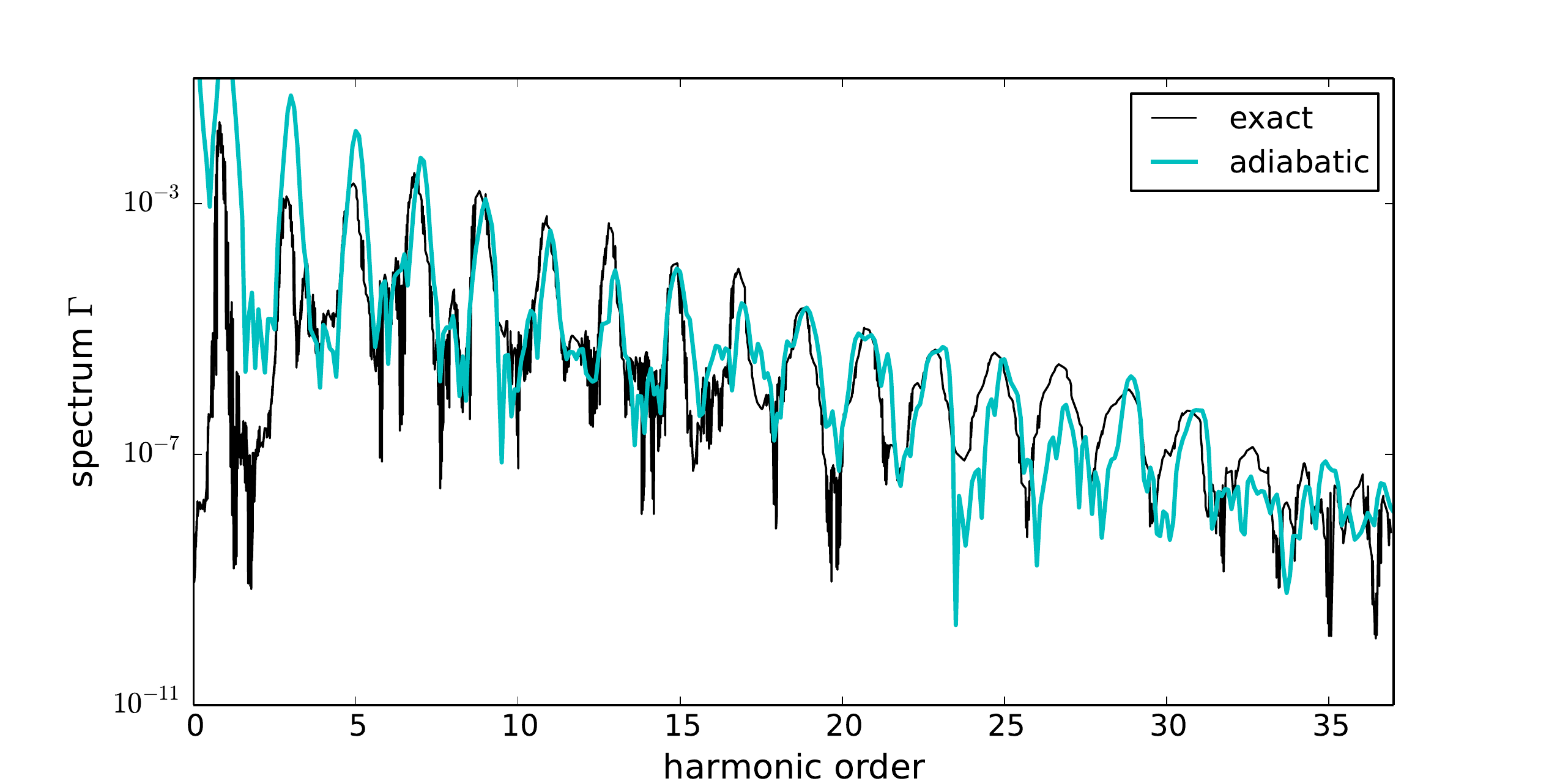}
  \caption{
  Logarithmic plot of the exact and the adiabatic electron dynamics 
  \eqref{eq:chias}, for the model of \cite{sukiasyan09a}.
  The exact spectrum was extracted graphically from FIG.\ 1 (a) of 
  \cite{sukiasyan09a}.}
  \label{fig:spectrum_prl}
\end{figure}

Finally, we also made a calculation for one of the best exact 2-electron models 
available in the literature:
The N$_2$ model of \cite{sukiasyan09a,sukiasyan10} for a system of two fixed 
nuclei and two electrons that may move each in two dimensions, in an intense 
10-cycle laser pulse.
Figure \ref{fig:spectrum_prl} shows how the spectrum of the AE approximation 
compares to the exact spectrum.
The harmonic intensities seem to be well reproduced for most harmonics, at 
least on the logarithmic scale of the figure.

In summary, we have shown that the HHG spectra can in principle be 
obtained from a single-electron Schr\"odinger.
A promising way to obtain the corresponding potential in practice is to make 
the adiabatic electron approximation, an extension of the concept of the 
Born-Oppenheimer approximation to systems of identical particles.
The main advantage of this approach is that only a single-electron 
Schr\"odinger equation needs to be solved, that all many-electron effects are 
included to a certain extent, and that it proposes a unique way of obtaining 
the required potential:
The energy of the system with one clamped negative charge needs to be 
determined for various positions of the clamped charge.
This problem is a standard, albeit challenging one in electronic structure 
theory.

Furthermore, our proposed approach can be systematically improved by 
the transfer of concepts of non-adiabatic molecular dynamics to systems of 
identical particles.
The large knowledge base in this field makes us confident that an appropriate 
approximation to the exact electron factorization can be found for many 
different molecular systems in strong electric laser fields.

\bibliography{hhg_final}{}

\end{document}